\documentclass[preprint,showpacs,showkeys,preprintnumbers,amsmath,amssymb]{revtex4}

\usepackage{graphicx}
\usepackage{dcolumn}
\usepackage{tabularx}
\usepackage{bm}
\usepackage{epsfig}
\usepackage{xcolor}
\usepackage{ucs}
\usepackage[utf8x]{inputenc}
\DeclareGraphicsExtensions{.ps}

\begin{document}

\title{Towards an efficient perovskite visible-light active photocatalyst}

\author{M.M. Fadlallah$^{1,2}$}\
\author{T.J.H. Vlugt$^{3}$}\
\author{M.F. Shibl$^{4,5}$}\
\affiliation{
$^1$Centre for Fundamental Physics, Zewail City of Science and Technology, Giza, Egypt\\
$^2$Physics Department, Faculty of Science, Benha University, Benha, Egypt\\
$^3$Engineering Thermodynamics, Process \& Energy Laboratory, Delft University of Technology, Leeghwaterstraat 39, 2628 CB Delft, The Netherlands\\
$^4$Gas Processing Center, College of Engineering, Qatar University, P.O. Box 2713, Doha, Qatar\\
$^5$Department of Chemistry, Faculty of Science, Cairo University, Cairo, Egypt
}

\date{\today}

\begin{abstract}
The band gap of SrTiO$_{3}$ retards its photocatalytic application. Regardless narrowing the band gap of the anion doped
SrTiO$_{3}$, the anion doping structures have low photoconversion efficiency. The co-cation dopings are used to modify the band gap and band edges positions
of SrTiO$_{3}$ to enhance the photocatalyitic properties by extending the absorption to longer visible-light. Using density 
functional calculations with the Heyd-Scuseria-Ernzerhof (HSE06) hybrid functional for exchange-correlation, the crystal and electronic structures as well as the optical properties of 
the mono and codped SrTiO$_{3}$ are investigated. The (Mo, Zn/Cd) and (Ta, Ga/In) codoped SrTiO$_{3}$ are explored in the way to retain the semiconductor characteristics of the latter. It is found that (Ta, Ga/In) codoping does not enhance the photocatalytic activity of SrTiO$_3$ due to its large band gap. Moreover, the position of the conduction band edge of the (Mo, Cd) monodoping impedes the photocatalytic efficiency. The obtained results indicate that (Mo, Zn) codoped SrTiO$_3$ 
can potentially improve the photocatalyitic activity.

\end{abstract}

%\pacs{73.20.-r, 73.40.Rw}


\maketitle

\section{Introduction}

Designing efficient materials to minimize or exclude utilizing hazardous substances is the heart of sustainable (green) life. Hydrogen production from solar water splitting as renewable green energy resources and minimizing environmental pollution by photoreduction of carbon dioxide via photocatalysis are few examples. Although photocatalysis efficiency has been extensively studied both experimentally and theoretically for decades, finding efficient photocatalyst is still the focus of many researchers. New course of metal-oxide materials has contributed to photocatalysis. However, most of the metal-oxide photocatalysts comprise large band gap causing lack of photoactivity under visible light. For instance, the ability of perovskites to absorb the visible spectrum is limited by their optical band gaps. Consequently, many experimental 
efforts to reduce perovskites optical gaps by chemical doping  have been made \cite{A1,A2,A3,A4}.

SrTiO$_{3}$ as a perovskite candidate is an oxide semicoductor with a band gap of $3.2$ eV at room temperature. SrTiO$_{3}$ can be 
considered as a promising photocatalyst because of its ability to split water into H$_{2}$ and O$_{2}$ without the application of an 
external electric field \cite{A5,A6,A7,A8}. The large band gap (3.2 eV) activates the photocatalytic properties of SrTiO$_{3}$ by 
absorbing only UV radiation, which represents $3$\%  of solar spectrum. In order to enhance the photocatalytic efficiency of SrTiO$_{3}$ 
in the visible light region, introducing doping states into the band gap and/or narrowing the latter are the common methods. Furthermore, 
the doping into a semiconductor can create a new optical absorption edge which is very important in the photocatalysis process. 

Many experimental and theoretical publications have been performed to develop the photocatalytic efficiency of SrTiO$_{3}$ via doping 
in either cationic sites (mostly Ti) or the oxygen anionic site. In the band structure of SrTiO$_{3}$, oxygen 2p-orbitals dominate in 
the valence band (VB) whereas Ti 3d-orbitals prevail in the conduction band (CB). Generally, anion doping serves mostly to modify the 
VB of SrTiO$_{3}$ due to different p-orbital energy of the dopant material, while cation doping usually produces gap states in the 
forbidden region or resonates with the bottom of the CB. For anionic dopant material, the N-doping has been tackled theoretically 
and experimentally. N-doping increases the photocatalytic activity by introducing mid-gap states which can 
prevent the charge carrier mobility resulting in decreasing the photoconversion efficiency \cite{A9,A10,A11,A12}. Moreover, C, F, P, S and B dopings
have been studied to improve the photocatalytic properties, it is found that C doping at the Ti site decreases the band gap and thus improving their photocatalytic  properties \cite {A12,A13,A14}. On the other hand, cation 
doping using 
transition metals can increase the visible light activity but cannot activate water splitting such as Cr \cite{A15,A16,A8}, Mn, Ru, Rh and 
Pd \cite{A2,A17,A18,A19}. Furthermore, anion-cation, anion-anion or cation-cation codoping pairs have been used to enhance 
the photoresponse \cite{A10,A17,A18,A19,A20}. However, to our knowledge, all the co-cation doping studies have been performed without giving attention to SrTiO$_{3}$ semiconductor characteristics.

In this contribution, designing a novel and efficient visible-light active photocatalyst (SrTiO$_{3}$) employing cation codoping is inspected taking into account the following criteria: the co-cation i) should 
not change the SrTiO$_{3}$ semiconductor characteristics, ii) does not make any significant distortion in the clean SrTiO$_{3}$ crystal, iii) reduces 
the band gap to absorb the visible light and iv) changes the band gap edge of the CB to fulfill the redox potential requirements for water splitting. Since the electronic 
configuration of Ti$^{4+}$ is [Ar] 3d$^{0}$4s$^{0}$, the most convenient cation dopant should have empty or completely filled valence d-orbital, i.e. d$^0$ or d$^{10}$ electronic 
configuration. Four co-cation dopings have been assumed to maximize the visible light activity of the photocatalyst that are (Mo$^{6+}$, Zn$^{2+}$/Cd$^{2+}$) and (Ta$^{5+}$, Ga$^{3+}$/In$^{3+}$).

The manuscript is organized as follows: in section \ref{calc}, the methods utilized in the calculations are represented. The pristine SrTiO$_3$ is discussed as a reference system for the rest of the calculations in section \ref{res}. Moreover, section \ref{res} outlines various cation monodoping collection which is the entry of different cation co-doping combinations 
emphasizing the band structure and its effects on the optical properties and photocatalytic activity.  Finally, section \ref{conc} concludes.

\section{Calculation Methods} \label{calc}

The spin polarized density functional theory (DFT) calculations are performed using the projector augmented wave (PAW) pseudopotentials in 
the Vienna \textit{ab initio} Simulations Package (VASP) code \cite{B1,B2}. For the exchange and correlation energy density functional, 
the generalized 
gradient approximation (GGA) \cite{B3,B4} in the scheme of Perdew-Bueke-Ernzerhof (PBE) \cite{B5} is utilized to get the optimzed structures. The PAW potentials 
with the valance electron 
Sr (4s$^{2}$ 4p$^{6}$ 5s$^{2}$), Ti (4s$^{2}$ 3d$^{2}$), O (2s$^{2}$ 2p$^{4}$), Mo (5s$^{2}$ 4d$^{4}$), Zn (4s$^{2}$ 3d$^{10}$), 
Cd (5s$^{2}$ 4d$^{10}$), Ta (6s$^{2}$ 5d$^{3}$), Ga (4s$^{2}$ 4p$^{1}$) and In (5s$^{2}$ 5p$^{1}$) are 
employed. The wave functions are expanded in plane waves up to cutoff energy of 600 eV. A Monkhorst-Pack $k$ point mesh \cite{B6} of 8 $\times$ 8 $\times$ 8 is used 
for geometry optimization until the largest force on the atoms becomes smaller than 0.01 eV/\AA\ and the tolerance of total energy reaches 10$^{-6}$ eV.
A 2 $\times$ 2 $\times$ 2 supercell containing 40 atoms is structured to simulate SrTiO$_{3}$. The stability, the electronic structures and the optical
properties are carried out using the hybrid 
functional Heyd-Scuseria-Ernzerhof (HSE06) \cite{B7,B8,B9}. The exchange-correlation energy in the hybrid functional (HSE06) is formulated as:
\begin{equation}
E^{HSE}_{xc} = 0.28E^{SR}_{x}(\mu)+(1-0.28)E^{PBE,SR}_{x}(\mu)+E^{PBE,LR}_{x}(\mu)+E^{PBE}_{c} ,
\end{equation}
The mixing coefficient (0.28) and the screening ($\mu=0.2 \AA^{-1}$) parameters are considered to get the closer to the experimental band gap
value \cite{B10,B11}. A 3 $\times$ 3 $\times$ 3 kpoint in the Brioullioun zone is utilized in the HSE06 calculations. Using the frequency dependent dielectric 
function as implemented in VASP, the optical properties are investigated. Figure 1 illustrates the crystal structure of pristine, mono-doped, and codoped SrTiO$_{3}$
which are used in this study

\begin{figure}[h]
\includegraphics[width=0.3\textwidth]{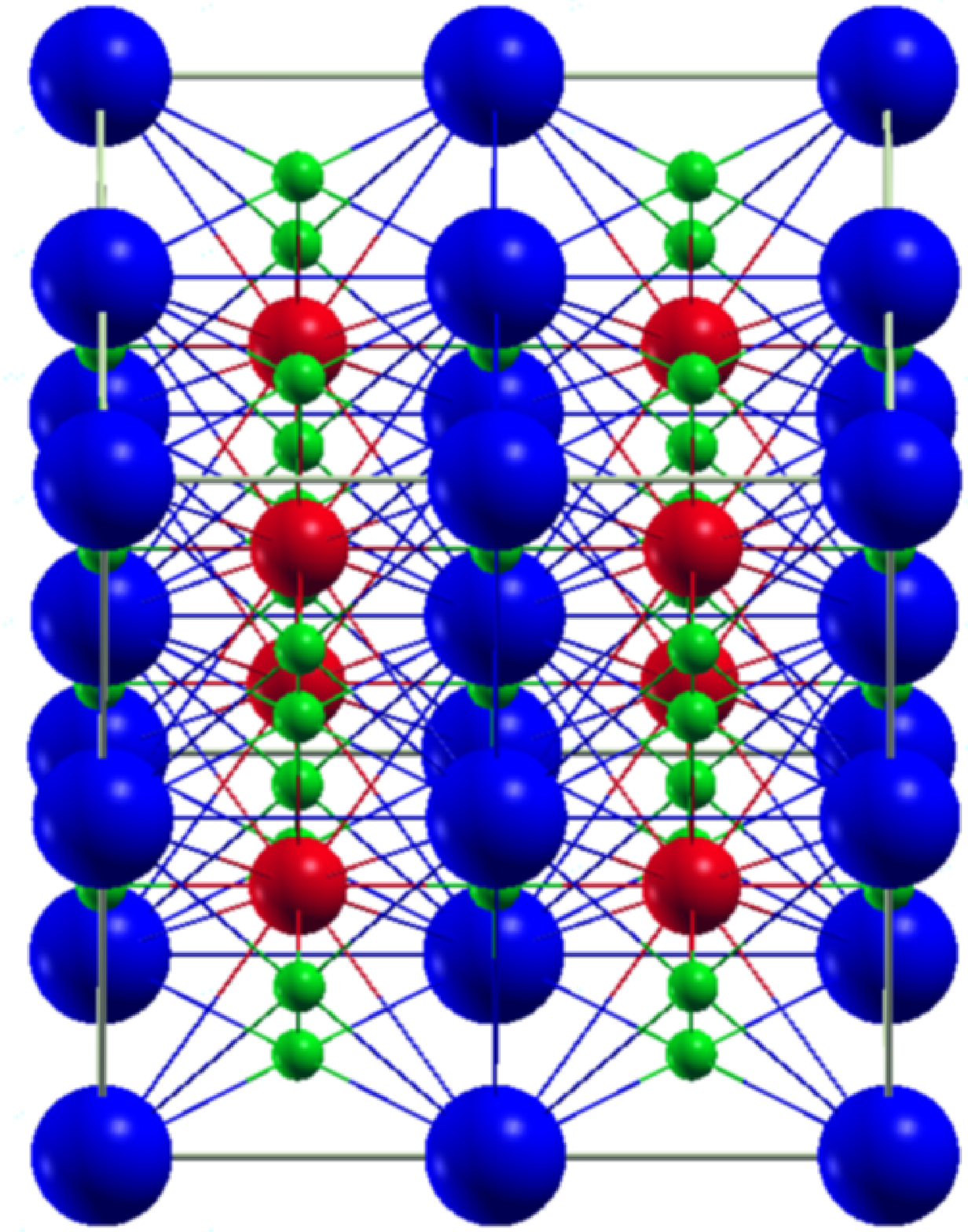}~\hspace{2mm}
\includegraphics[width=0.3\textwidth]{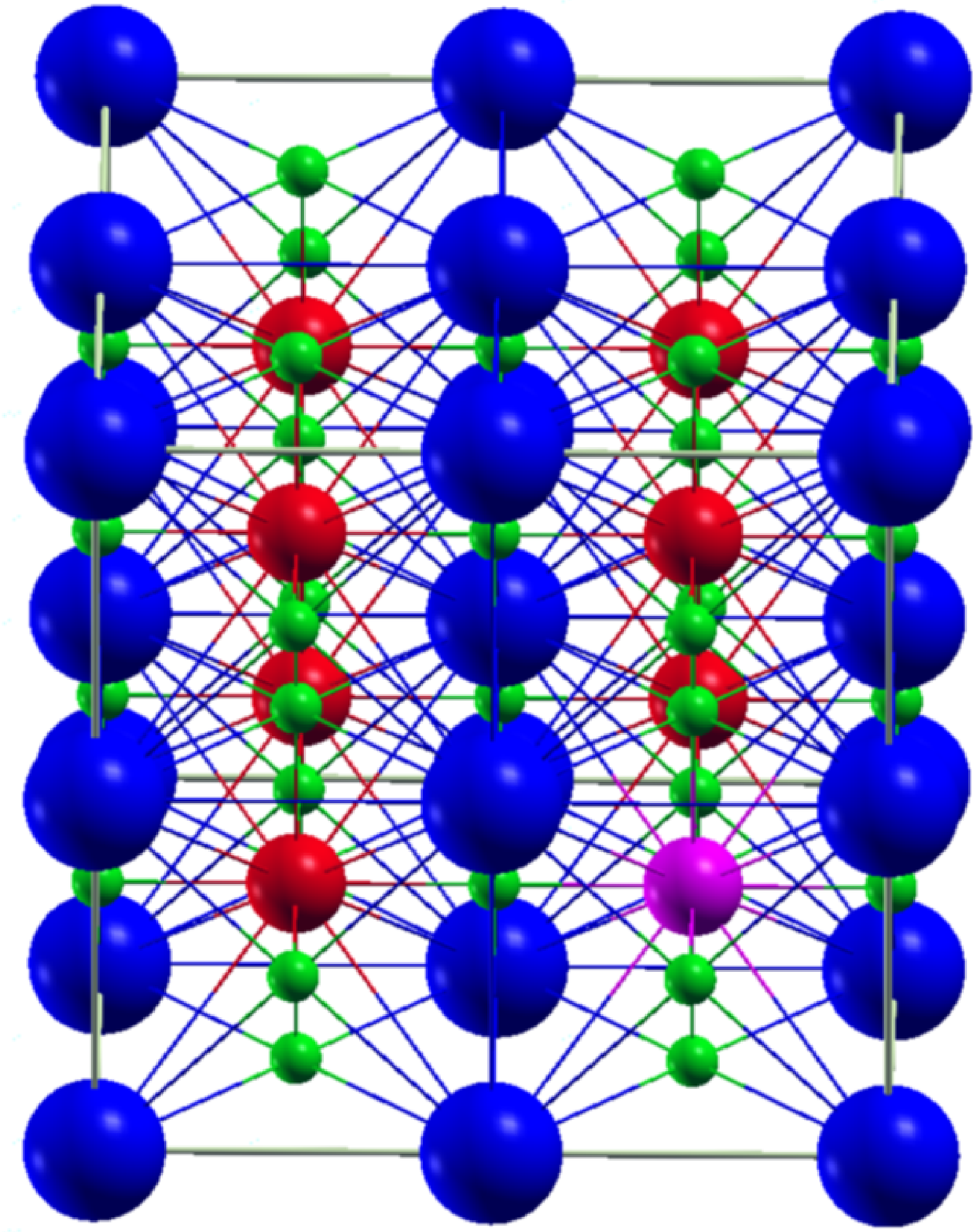}~\hspace{2mm}
\includegraphics[width=0.3\textwidth]{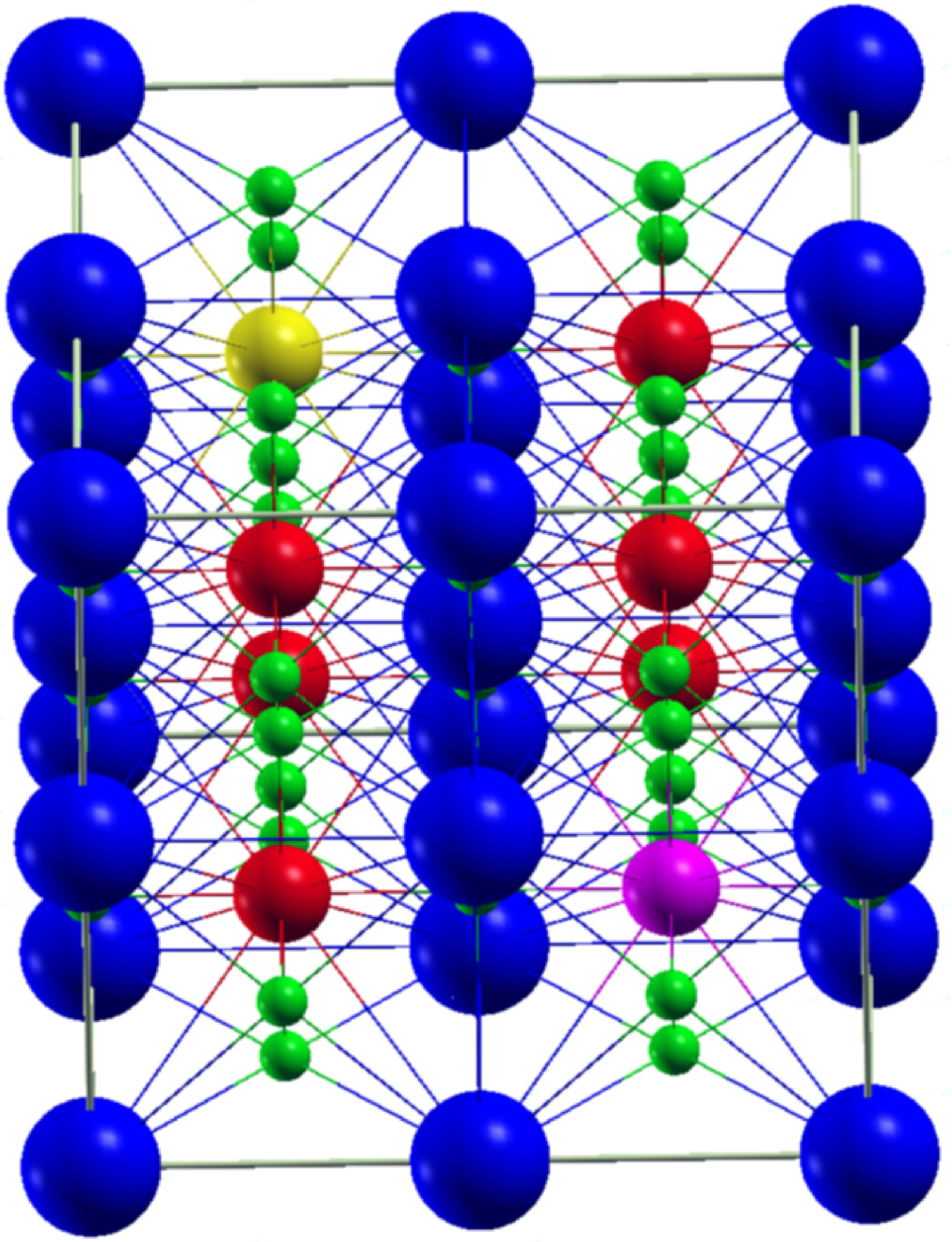}
\caption{(Color online) 2 x 2 x 2 supercell for pristine, mono-doped, and codoped SrTiO$_{3}$ crystal structures. 
Blue, red, green, pink, and yellow spheres represent Sr, Ti, O, first mono (M1) doped, and second doped (M2) atoms, respectively.}
\label{fig1}
\end{figure}

\section{Results and discussion} \label{res}

\subsection{Pristine SrTiO$_{3}$}

In order to study the effect of dopants on the electronic structure of SrTiO$_{3}$, first the pristine SrTiO$_{3}$ is addressed.
The relaxed Sr$-$O and Ti$-$O bond lengths are 2.755 \AA\ and 1.945 \AA\, respectively, which are in good agreement with previous 
experimental results and theoretical calculations \cite{A1,A4}.
The density of states (DOS) and projected density of states (PDOS) of pristine structure are depicted in Figure 2. The top of valance band is dominated by O states and the 
contribution of Ti states can be considered below $-3$ eV  which refers to the covalent bond Ti$-$O. However Ti states dominate in the bottom 
of conduction band and O states have good contributions in the whole range of conduction band. A small density of Sr states is observed in the whole energy range which reveals ionic interaction
between Sr and TiO$_{6}$ octahedron. The band gap is found to be $3.2$ eV which is in line with the experimental values \cite{A6}.

\begin{figure}[h]
\includegraphics[width=0.5\textwidth]{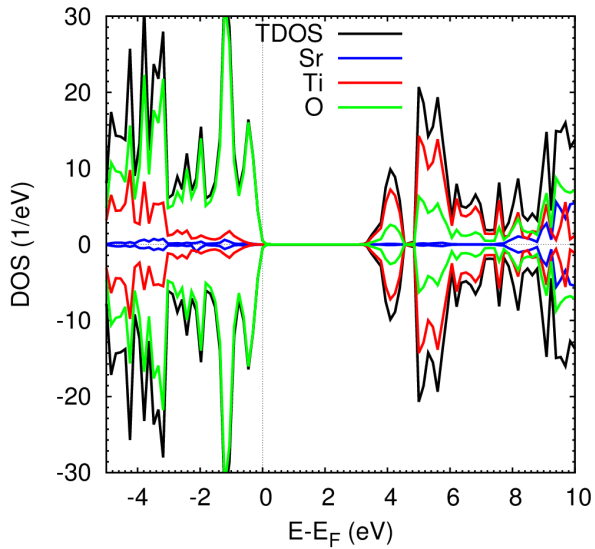}~\hspace{2mm}
\caption{(Color online) Density of states of pristine SrTiO$_{3}$.}
\label{fig2}
\end{figure}

\subsection{mono-doped SrTiO$_{3}$}

{\bf Mo/Ta doping} \\

It is useful to discuss the influence of mono-doping on the SrTiO$_{3}$ electronic structure to understand the co-doping effect. 
First a Ti atom is replaced by a Mo one in the pristine SrTiO$_{3}$. The optimized structure shows that the Mo$-$O bond 
length ($1.944$ \AA) does not change significantly as compared to Ti$-$O in pristine crystal due to comparable ionic sizes of Mo$^{6+}$ ($R=0.60$ \AA) and Ti$^{4+}$ ($R=0.61$ \AA) \cite{C0,A3,C01}. 
To determine the stability of the mono-doped structures, the defect formation energy ($E_{f}$) has been calculated using 
the following expression:

\begin{equation}
E_{f} = E_{M-SrTiO_{3}}+\mu_{Ti}-(E_{SrTiO_{3}}+\mu_{M}),
\end{equation}
\
where $E_{M-SrTiO_{3}}$ and  $E_{SrTiO_{3}}$ are the total energies of metal-doped SrTiO$_{3}$ and pristine SrTiO$_{3}$, respectively. 
The $\mu_{Ti}$ and $\mu_{M}$ stand for the chemical potential for Ti and dopant atom, respectively, which are assumed as the energy of 
one metal atom in their corresponding metal bulk structure \cite{C011}. The  defect formation energy for Mo doping is found to be $3.36$ eV.
Figure 3 (left) displays the calculated density of states (DOS) and the projected one of the Mo-doped SrTiO$_{3}$, the band gap becomes $2.0$ eV 
which is less than that of the pristine structure ($3.2$ eV). The Fermi energy is shifted towards the edge of the CB suggesting an n-type
conducting character. This is due to the excess two electrons compiled by Mo atom (Mo is in its 6+ oxidation state). 
The magnetic moment of the Mo-doped SrTiO$_{3}$ becomes $2$ $\mu_{B}$. The mid gap states are created at $-0.8$ eV below the conduction band. The band 
structure is shifted towards the low energy 
compared to the pristine structure and the created states arise from the contribution of the localized Mo states.

\begin{figure}[h]
\includegraphics[width=0.5\textwidth]{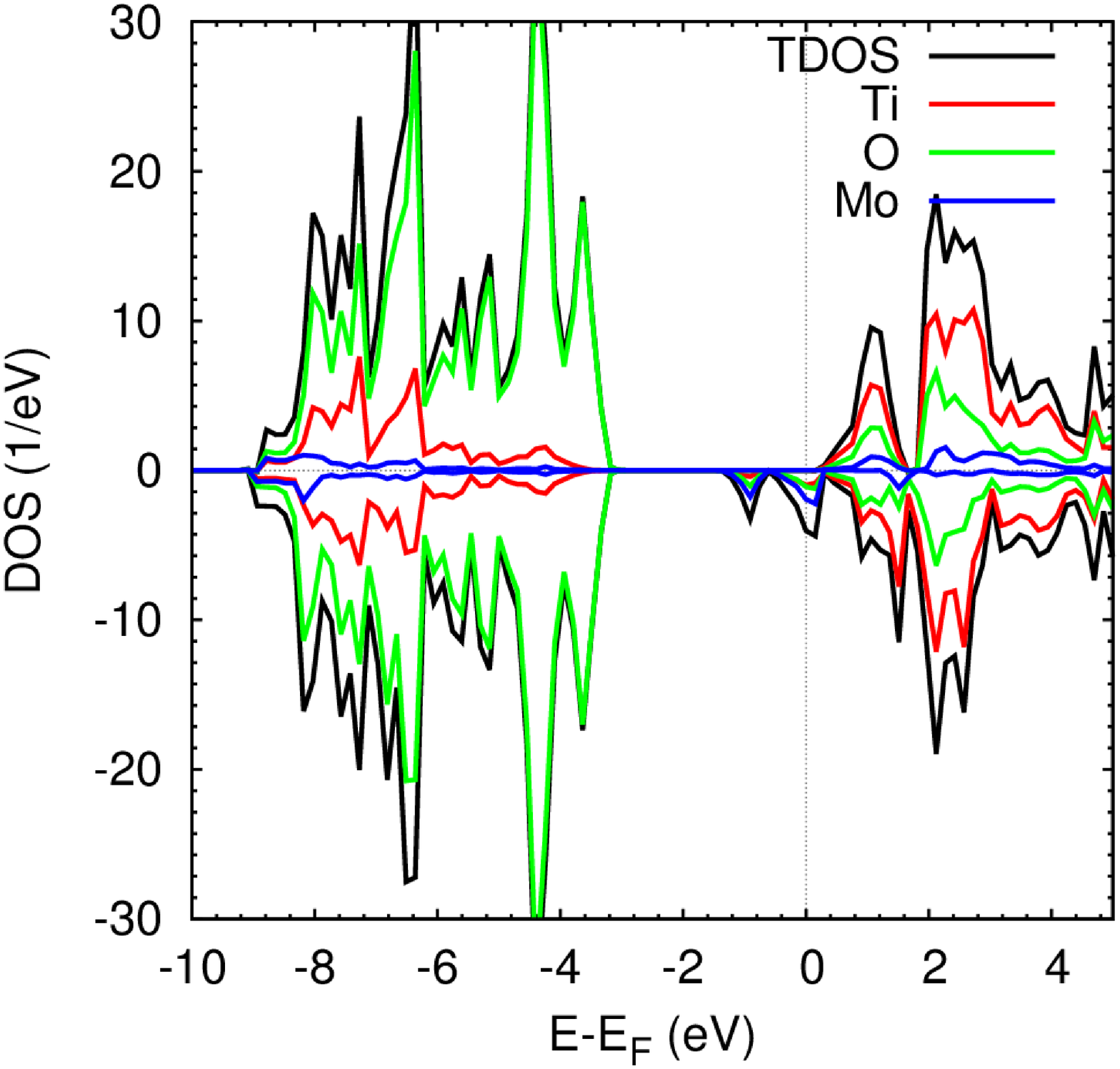}~\hspace{5mm}
\includegraphics[width=0.5\textwidth]{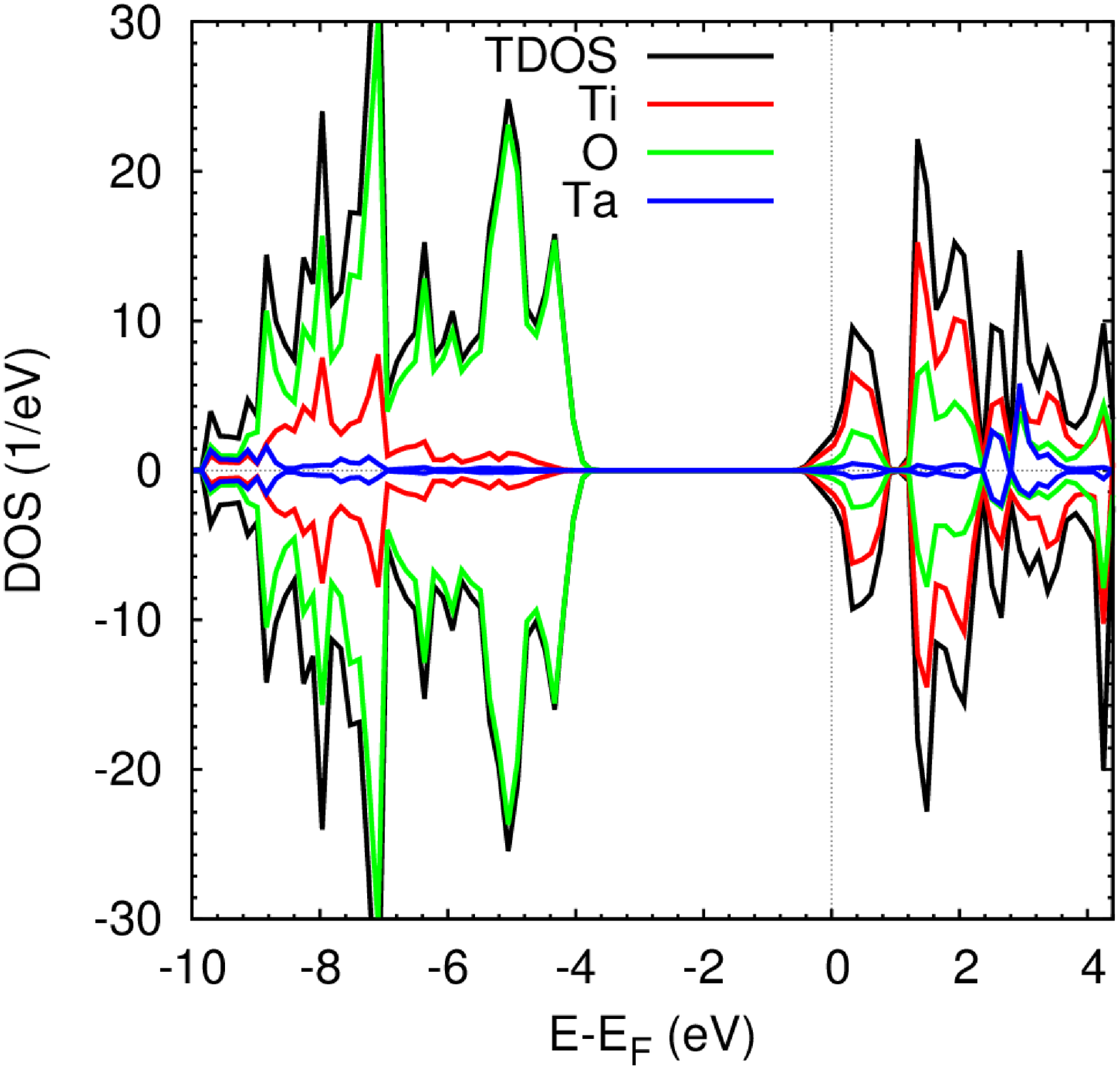}
\caption{(Color online) Density of states of (left) Mo- and (right) Ta-doped SrTiO$_{3}$.}
\label{fig3}
\end{figure}

For Ta-doped structure, the bond length of Ta$-$O becomes ($1.95$ \AA) which is slightly 
larger than that of Ti$-$O due to the comparable ionic sizes of Ta$^{5+}$ ($R=0.64$ \AA) and Ti$^{4+}$. The calculated defect 
formation energy using eq. (2) 
for Ta-doping gives $-3.21$ eV suggesting a more stable and lower energy cost than Mo-doped structures.
The effect of Ta-doping on the electronic structure is illustrated in Figure 3 (right). The Fermi level located in the CB indicates 
an n-type conducting behavior
(similar to Mo doping) due to the 5+ oxidation state of Ta. The band gap does not change as compared to the pristine 
structure. The low contribution of 
Ta-localized states appears in the CB but the O and Ti ones still prevail as seen in the pristine SrTiO$_{3}$. Moreover, like Mo-doping, 
the band structure is moved towards the low energy.

{\bf Zn/Cd doping} \\

Since, the ionic size of Zn$^{2+}$ ($R=0.7$ \AA) is slightly larger than that of Ti, there is a little 
local distortion of the crystal structure where the bond length of Zn$-$O ($R=1.982$ \AA) becomes larger than that of Ti$-$O. The 
local distortion and the magnetic moment (2 $\mu_{B}$) of the structure is related to the oxidation state 
of Zn$^{2+}$. The defect formation energy in this case is $12.9$ eV indicating less stablility of Zn-doped structure than Ta and Mo-doped 
ones. The localized Zn states contribute at the top of the valance band and the band gap becomes less than 
the pristine structure by $0.8$ eV as exhibited in Figure 4 (left). The Zn doping can experimentally improve the photocatalytic 
properties \cite{C1,C2}.
In the case of Cd-doped SrTiO$_{3}$,  the ionic size of Cd$^{2+}$ ($R=0.95$ \AA) is larger than that of Ti$^{4+}$, 
which reflects the increasing in the bond length of Cd$-$O ($2.034$ \AA) compared to the bond length of Ti$-$O in the 
pristine structure. Cd$^{2+}$ causes similar effect on the magnetic moment and distortion like Zn$^{2+}$ does. The defect 
formation energy becomes $15.48$ eV. The contribution of Cd-states appears at the top of valance band with O states. This contribution of 
Cd-dopant material reduces the band gap to $1.8$ eV, see Figure 4 (right). 

\begin{figure}[h]
\includegraphics[width=0.5\textwidth]{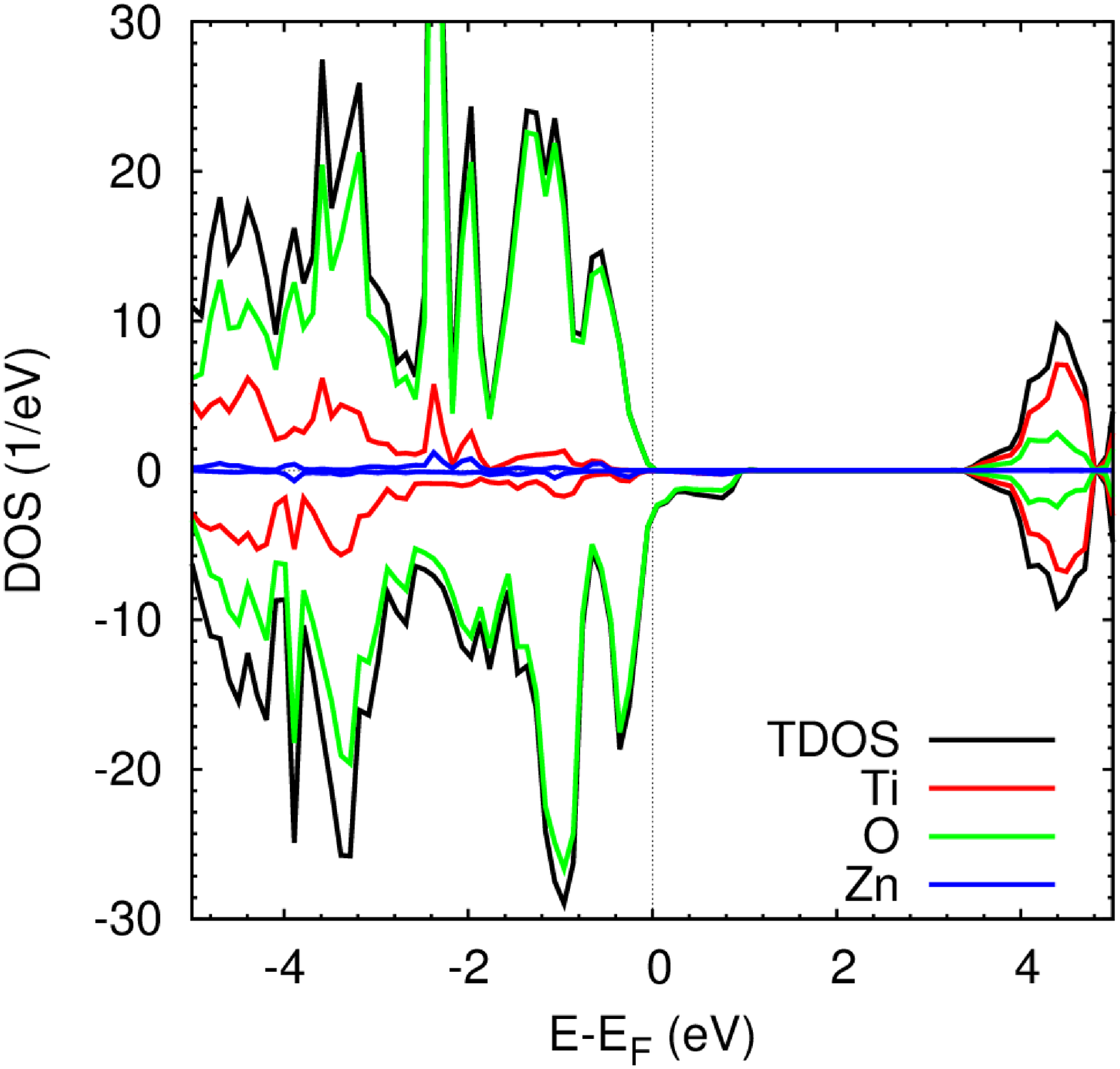}~\hspace{5mm}
\includegraphics[width=0.5\textwidth]{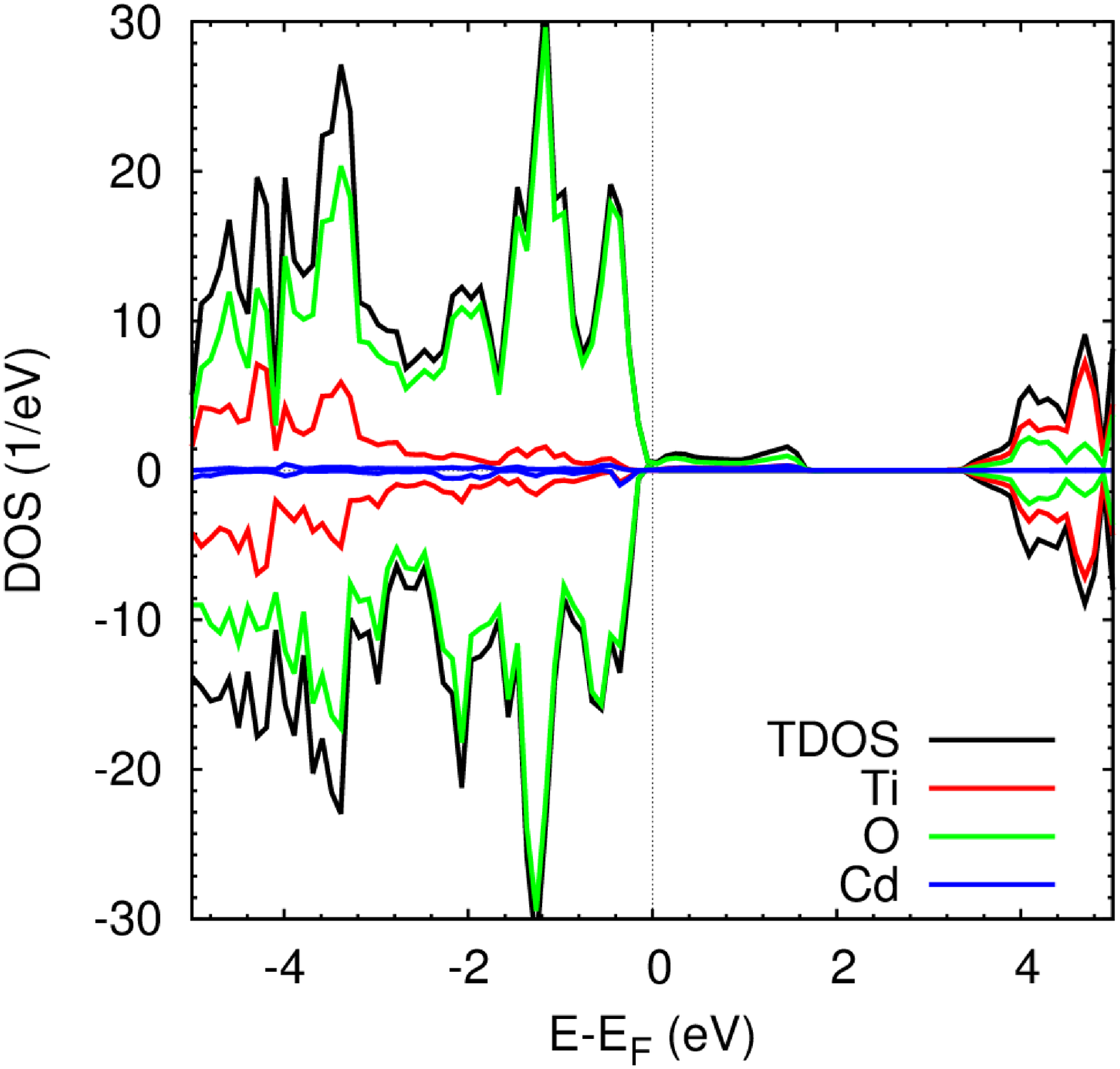}
\caption{(Color online) Density of states of (left) Zn- and (right) Cd-doped SrTiO$_{3}$.}
\label{fig4}
\end{figure}

{\bf In/Ga doping} \\

The last two single dopant atoms are In and Ga. 
The ionic size of In$^{3+}$ ($R=0.82$ \AA) is larger than that of Ti$^{4+}$. This size effect is manifested in the defect 
formation energy value 
($10.58$ eV) which is larger than Ta-doped SrTiO$_{3}$. Furthermore, the bond length of In$-$O 
becomes $2.055$ \AA. The magnetic moment is 1 $\mu_{B}$ due to 3+ oxidation state of In. The influence of In-dopant on the 
electronic structure of SrTiO$_{3}$ is shown in Figure 5 (right). The trivalent oxidation state for In creates a vacancy in the system and 
analysing the PDOS indicates that the In-states have insignificant contributions. The O states at the top of the valance band 
is disturbed along the spin down component and the band gap does not change compared to the pristine SrTiO$_{3}$.
Regarding Ga-doped SrTiO$_{3}$, Ga$^{3+}$ has ionic size ($0.62$ \AA) which is slightly larger than Ti$^{4+}$. Hence, the Ga$-$O bond 
length ($R=1.971$ \AA) is slightly larger than Ti$-$O and the defect formation energy becomes $7.48$ eV. The magnetic moment and the 
band gap of Ga-doped SrTiO$_{3}$ is 
very similar to In-doped one, see Figure 5 (left).

\begin{figure}[h]
\includegraphics[width=0.5\textwidth]{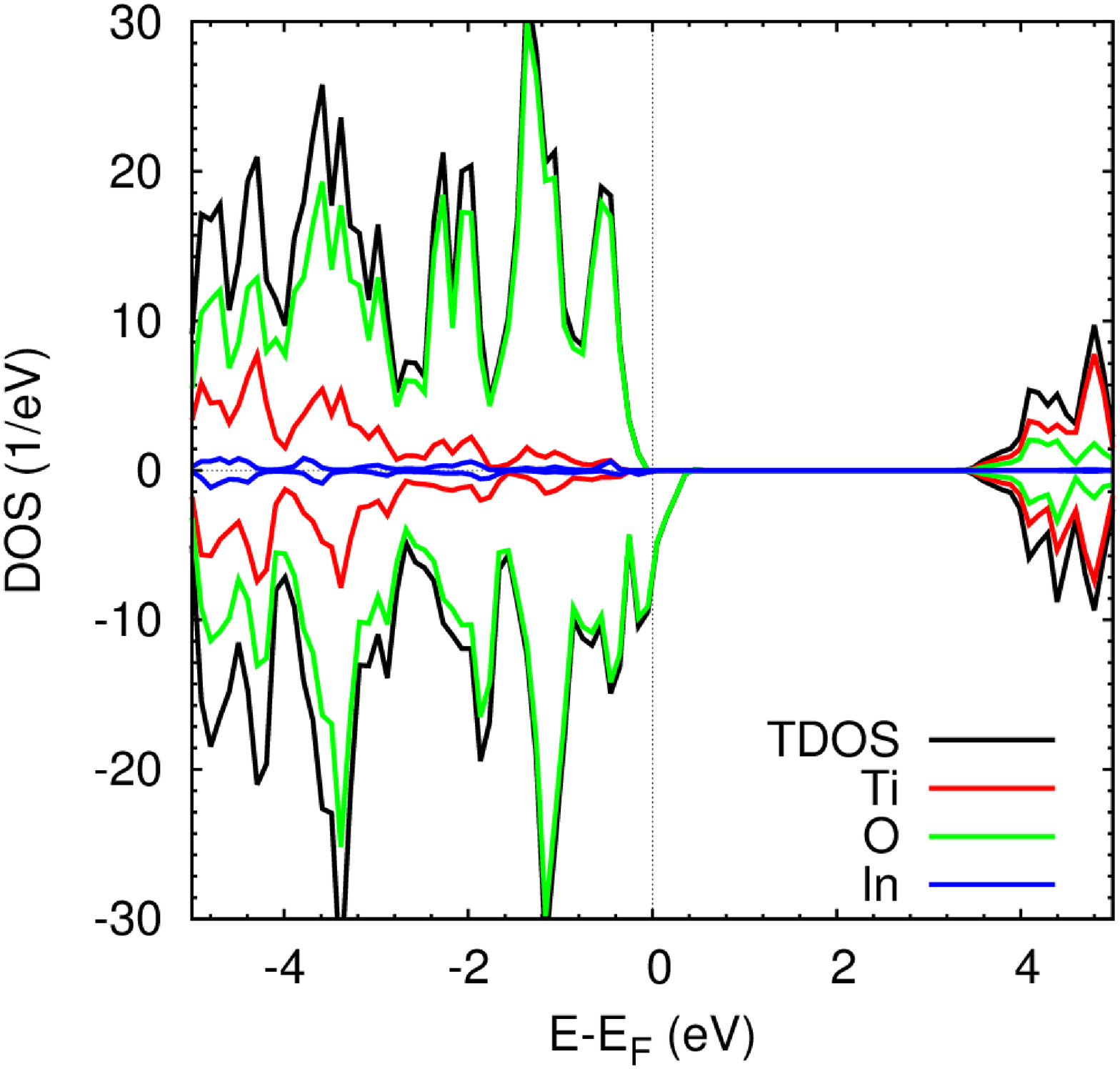}~\hspace{5mm}
\includegraphics[width=0.5\textwidth]{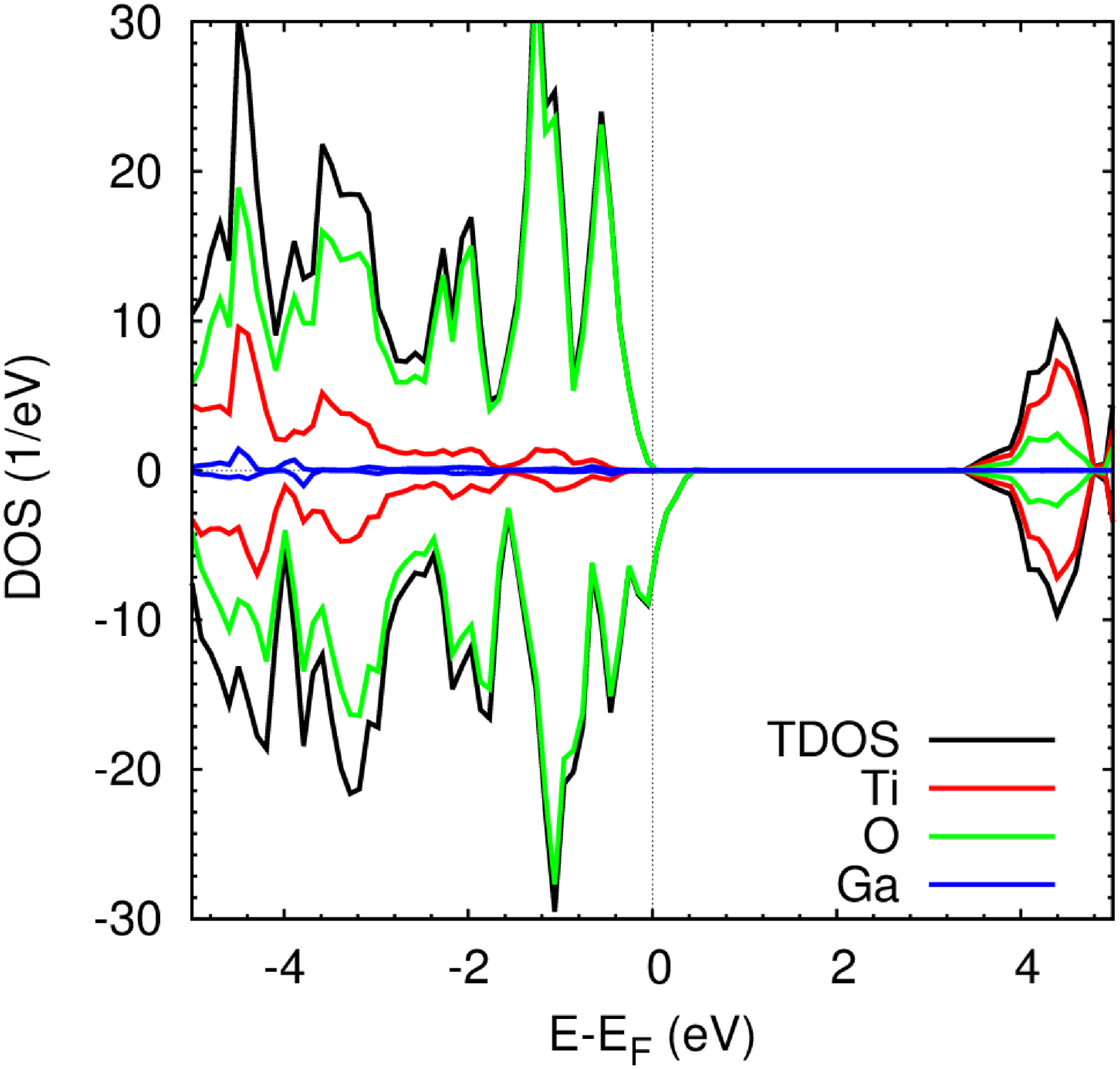}
\caption{(Color online) Density of states of (left) In- and (right) Ga-doped SrTiO$_{3}$.}
\label{fig5}
\end{figure}

\subsection{Codoped SrTiO$_{3}$}

{\bf (Mo, Zn/Cd) codoping} \\

So far, the mono-doped SrTiO$_{3}$ has been discussed in a bit details but the question now is whether the co-doping improves the 
properties of the pristine material. First, one can substantially look at the geometry and stability of the codoped material. In the
case of (Mo/Zn) co-doping, both Mo and Zn dopants change the bond lengths of Mo$-$O and Zn$-$O as compared to the corresponding ones in 
Mo and Zn mono-doped SrTiO$_{3}$. The Mo$-$O becomes shorter ($1.903$ \AA) while Zn$-$O turns out to be longer ($2.012$ \AA). 
The corresponding defect formation energy of the codoped SrTiO$_{3}$ is given by: 

\begin{equation}
E_{f} = E_{M1,M2-SrTiO_{3}}+2\mu_{Ti}-(E_{SrTiO_{3}}+\mu_{M1}+\mu_{M2}),
\end{equation}

where $E_{M1,M2-SrTiO_{3}}$ is the energy of the codoped SrTiO$_{3}$ structure,  $\mu_{M1}$ and $\mu_{M1}$ represent the chemical potential for first and second metals, respectively.
The defect formation energy of the (Mo, Zn) codoped, $12.4$ eV, is very close to that of 
Zn-doped SrTiO$_{3}$. Moreover, the stability of the codoped structure can be determined using the defect pair binding energy which 
is calculated by \cite{C03,C031}:

\begin{equation}
E_{b} = E_{M1-SrTiO_{3}}+E_{M2-SrTiO_{3}}+2\mu_{Ti}-(E_{M1,M2-SrTiO_{3}}+E_{SrTiO_{3}}),
\end{equation}

where $E_{M1-SrTiO_{3}}$ and $E_{M2-SrTiO_{3}}$ are the energies of the mono-doped first and second metals, respectively. The defect pair binding energy 
is calculated as $E_{b}=4.45$ eV where the positive value indicates that the codoped structure 
is sufficiently stable \cite{A1}. 

Second, it is necessary to explore how the electronic structure of SrTiO$_{3}$ changes upon (Mo, Zn) co-doping. 
The Fermi energy is located right above the valance band, similar to the pristine structure. The band gap decreases significantly 
to $1.8$ eV as compared to 
the pristine SrTiO$_{3}$, which allows the structure to absorb visible light. Furthermore, the band gap is similar to that of 
Mo-mono-doped SrTiO$_{3}$. The top of the valance band consists of O and Zn states with a little contribution of Ti and Mo states. However, 
Mo dominates 
in the bottom of the conduction band with the O states. (Mo, Zn) codoped material reduces the charge carrier loss and forms a charge 
compensated (n-p compensated) without emerging gap states or splitting the valance band.

\begin{figure}[h]
\includegraphics[width=0.5\textwidth]{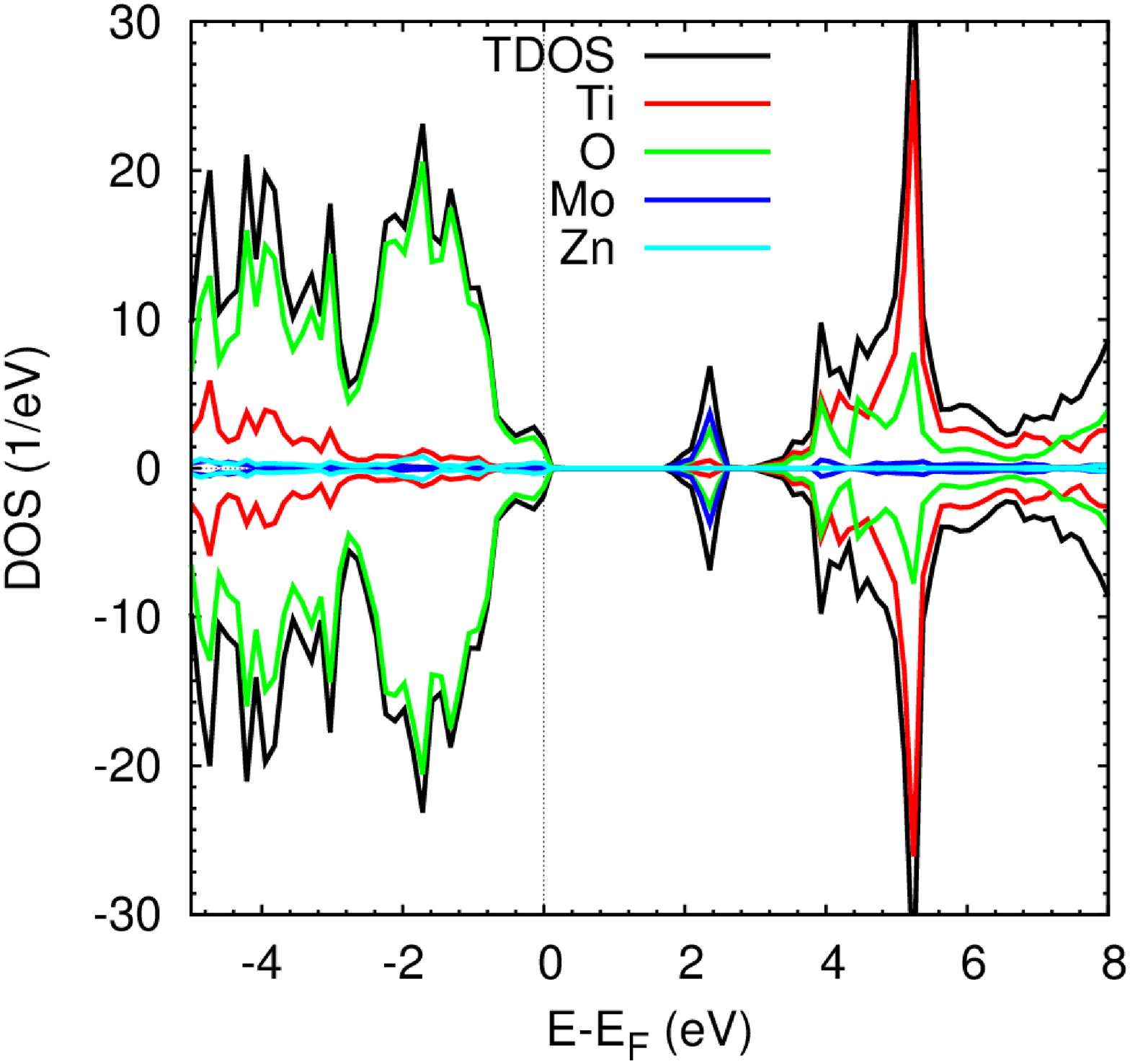}~\hspace{5mm}
\includegraphics[width=0.5\textwidth]{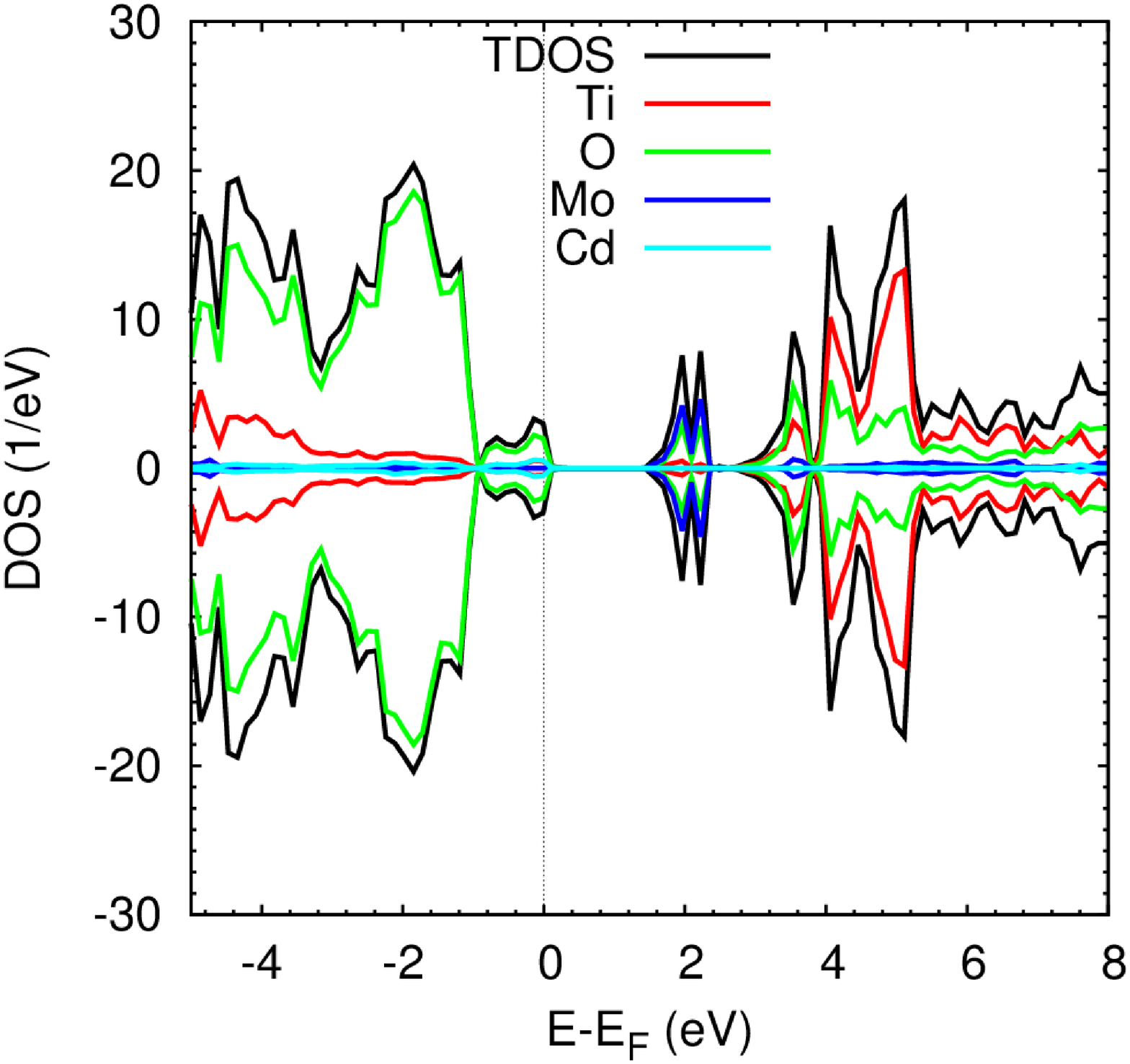}
\caption{(Color online) Density of states of (left) (Mo, Zn)- and (right) (Mo, Cd)-codoped SrTiO$_{3}$.}
\label{fig6}
\end{figure}

For (Mo, Cd) codoped SrTiO$_{3}$, the bond length Mo$-$O in the codoped structure ($1.923$ \AA) is shorter than the corresponding 
one in the Mo-mono-doped structure. Whereas, the Cd$-$O bond length in the codoped 
structure ($2.134$ \AA) is longer than that in the Cd-mono-doped structure. The formation energy for (Mo, Cd) codoped, $14.94$ eV, is higher 
than 
that of (Mo, Zn) codoped SrTiO$_{3}$ due to the large value of the formation energy of Cd. However the defect binding energy ($4.18$ eV) 
is lower than the corresponding one of the (Mo, Zn) co-doping.
Figure 6 (right) reveals the density of states and the projected ones of the (Mo, Cd) codoped SrTiO$_{3}$. The band gap becomes $1.6$ eV. 
Cd and Ti states contribute at the top of the valance band with dominated O states. Mo and O states have good contributions 
at the bottom of the conduction band with a lower contribution of the Ti states on contrast to the pristine junction where the latter 
contribute at the top of 
conduction band. \\

{\bf (Ta, In/Ga) codoping} \\

Last but not least, (Ta, In/Ga) co-doping is explored in this subsection. The bond lengths of Ta$-$O and In$-$O 
are $1.972$ \AA and $2.085$ \AA, respectively, which are slightly longer than the corresponding 
bond length in Ta and In mono-doped SrTiO$_{3}$. The defect formation and pair binding energies are $3.84$ eV 
which are less than of (Mo, Zn/Cd) codoped SrTiO$_{3}$. The band gap ($3.5$ eV) becomes larger than the pristine structure. The In 
states contribute at 
the top of the valance band similar to Ti states but O states still dominate. The bottom of the conduction band consists of analogous 
contribution of the Ti and O states but lower contribution of the Ta states.

\begin{figure}[h]
\includegraphics[width=0.5\textwidth]{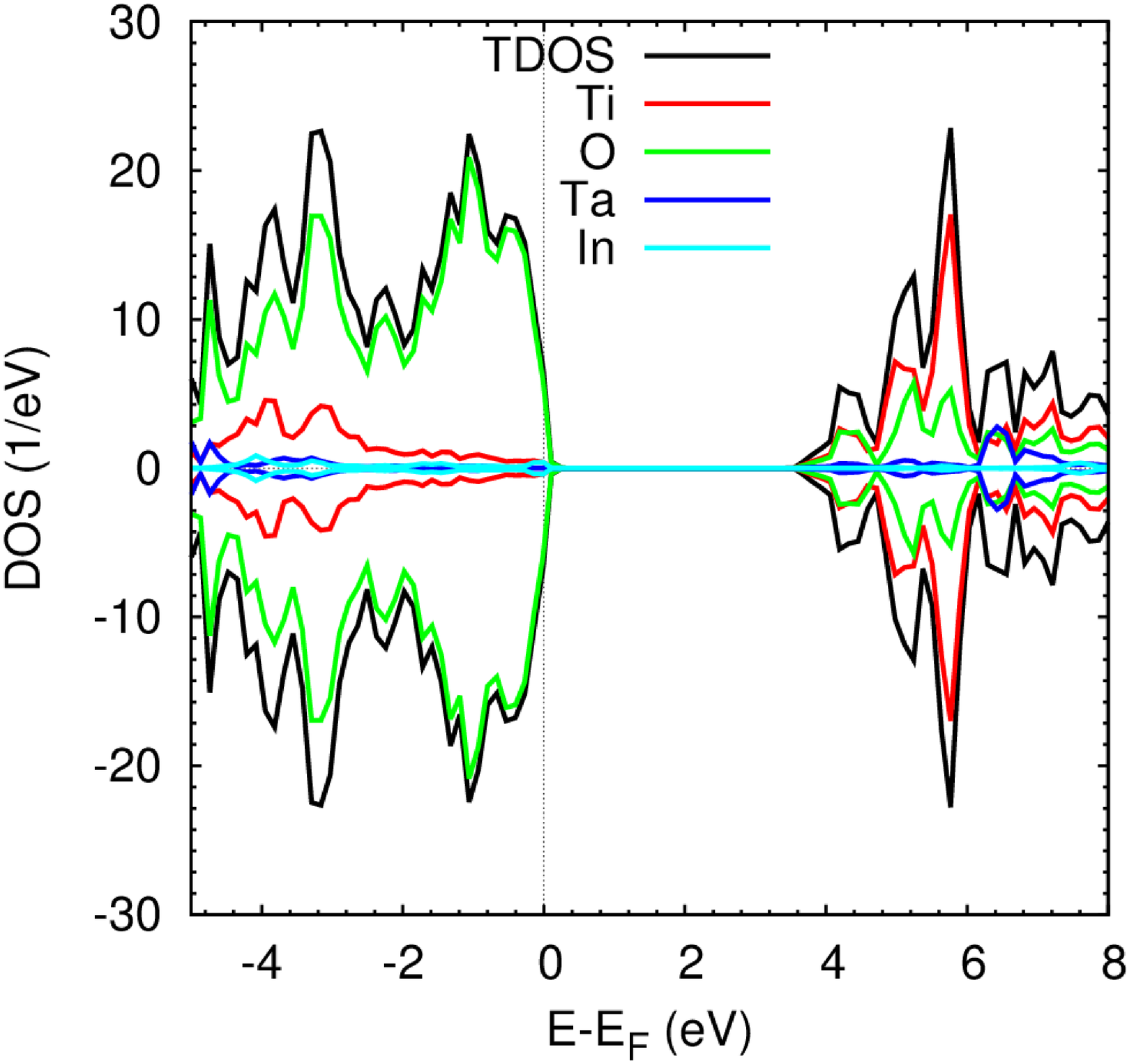}~\hspace{5mm}
\includegraphics[width=0.5\textwidth]{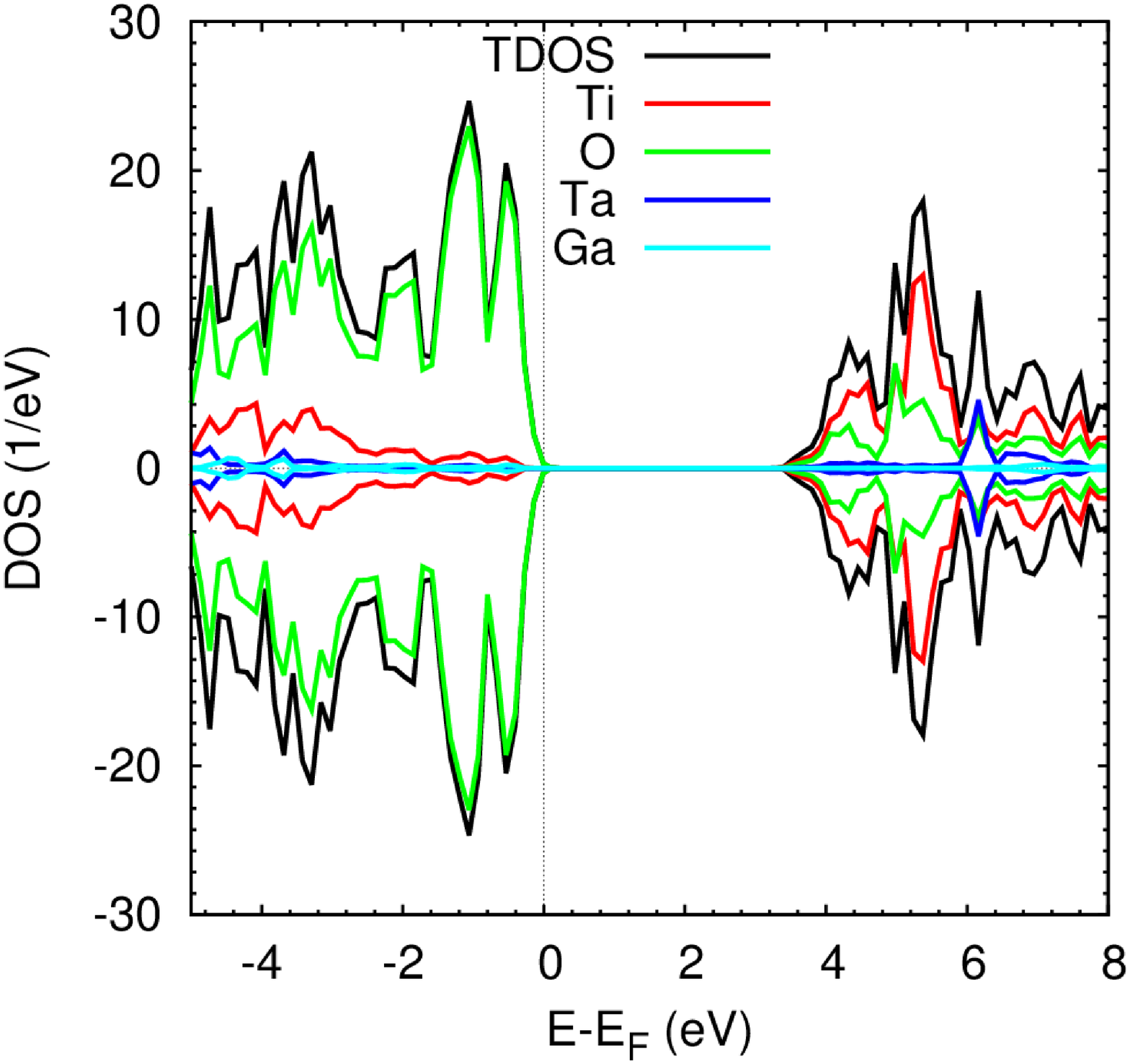}
\caption{(Color online) Density of states of (left) (Ta, In)- and (right) (Ta, Ga)-codoped SrTiO$_{3}$.}
\label{fig2}
\end{figure}

The bond lengths of Ta$-$O and Ga$-$O are $ 1.97$ \AA~which are a bit longer than the corresponding bond lengths in Ta and 
In mono-doped SrTiO$_{3}$. The defect formation energy in this structure ($7.65$ eV) is higher than (Ta, In) codoped analogue. However the 
defect pair 
binding energy ($3.82$ eV) is lower than the (Ta, In) codoped system. The DOS of (Ta, Ga) codoped SrTiO$_{3}$ shows that the band gap becomes 
($3.5$ eV) in line with that of (Ta, In) codoped system. The contribution of the Ga states at the top of valance band is small compared 
to Ti and O states. On the other hand, Ta contributes little to the bottom of the conduction band 
compared to Ti and O states. Ti states dominate in contrast to (Ta, In) codoped structure. Further, the band gap is larger than the pristine structure. 
Compared to the pristine structure, (Ta, In) and (Ta, Ga) co-doping cannot be used to extend the absorption edge to the 
longer-wavelength visible light. Fortunately, 
the gaps of the (Mo, Zn) and (Mo, Cd) codoped are smaller than the pristine SrTiO$_{3}$. Zn 3d and Cd 4d are located at the top of 
the valance band and the Mo states prevail at the bottom of the former. Due to the movement of the conduction band bottom to 
the lower energy, the band gap tightens without any gap states. The mid-gap or localized states
are not created due to the n-p compensated co-doping in all studied structures. 
Therefore, Mo together with Zn or Cd represent perfect co-doping candidates to enhance the photocatalytic properties. 
Since the ionic size plays an important role from the experimental point of view, the Mo$^{6+}$ and Zn$^{2+}$ dopants which are 
of comparable ionic radii to that of Ti$^{4+}$ suggest appropriate co-doping materials. Notice that, the effect of dopant concentration on the electronic structure is done using 2 $\times$ 2 $\times$ 1
and 2 $\times$ 2 $\times$ 3 supercells. It is found that the behavior of DOS does not change as compared to that of 2 $\times$ 2 $\times$ 2 supercell in line with the effect of different dopant concentrations on SrTiO$_{3}$  \cite{A1,A3}. 

\subsection{Optical properties}

The optical absorption properties of a semiconductor photocatalyst is to a large extent linked with its electronic structure which, in turn, affect the photocatalytic activity \cite{C3}.
Absorption in the visible light range is crucial for improving the photocatalytic activity of the doped SrTiO$_{3}$. The optical 
properties are determined by the angular frequency ($\omega$) dependent complex dielectric function
$\varepsilon_{1}(\omega)=\varepsilon_{1}(\omega)+i\varepsilon_{2}(\omega)$, which depends solely on 
the electronic structure. The imaginary part of the dielectric function, $\varepsilon_{2}(\omega)$, can be calculated from the momentum matrix elements 
between the occupied and unoccupied states. In addition, the real part can be calculated from the imaginary part by the Kramer-Kronig relationship. 
The corresponding absorption spectrum, was evaluated as implemented in VASP \cite{A4,C4}:

\begin{equation}
\alpha(\omega) = \sqrt{2}\omega\left(\sqrt{\varepsilon_{1}^{2}(\omega)+\varepsilon_{2}^{2}(\omega)}-\varepsilon_{1}^{2}(\omega)\right)^{1/2},
\end{equation}

\begin{figure}[h]
\includegraphics[width=0.4\textwidth]{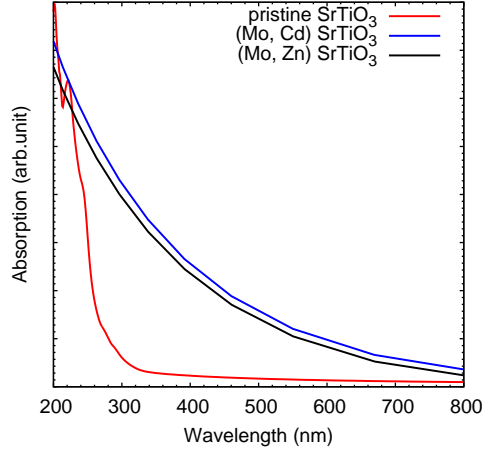}~\hspace{3mm}
\caption{(Color online) The absortion coefficient of the codoped SrTiO$_{3}$.}
\label{fig2}
\end{figure}

where $\alpha$ is the optical absorption coefficient. It can be seen that the pristine SrTiO$_{3}$ can 
only absorb the narrow UV light (360 nm) and shows no absorption activity in the visible light region, Figure 8. 
The calculated optical absorption spectra for (Mo, Cd) and  (Mo, Zn) codoped SrTiO$_{3}$ show absorption activity at the visible-light region in 
the range (450-800 nm). The optical absorption of compensated (Mo, Cd) codoped SrTiO$_{3}$ exhibits much more favorable visible-light 
absorption than compensated (Mo,Zn) codoped system, these results agrees with the earlier discussion

\subsection{Photocatalytic activity}
\begin{figure}[h]
	\includegraphics[width=0.6\textwidth]{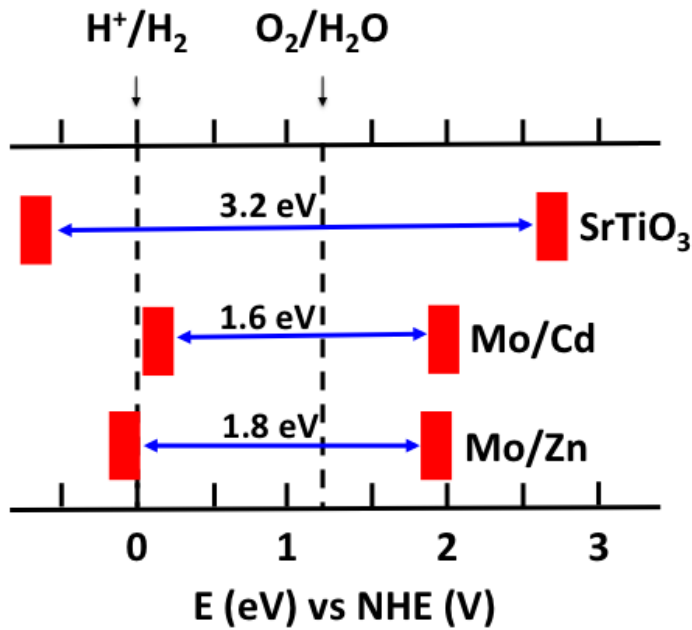}~\hspace{3mm}
	\caption{(Color online) The calculated band edges positions of the pristine and codoped SrTiO$_{3}$.}
	\label{ws}
\end{figure}

In this section, the desired thermodynamic conditions for the water splitting photocatalyst is discussed. The most influential criterion for 
the water splitting photocatalysis is the position of the relative band edges, i.e. the VB edge should lie below the H$_2$O/O$_2$ oxidation level and the CB edge is located above the H$^+$/H$_2$ reduction level. Assuming fixed VB edge position, the closer the CB edge to the  H$^+$/H$_2$ reduction level is, the better the photocatalytic efficiency becomes as this reduces the band gap.
Figure \ref{ws} shows that the calculated band edges positions of the pristine SrTiO$_{3}$ reveal water splitting photocatalytic activity. However, the large band gap does not promote the visible light absorption. 
In the case of (Mo, Cd) and (Mo, Zn) 
codoped SrTiO$_{3}$, The valance band edge for both codoped SrTiO$_{3}$ are located below the H$_2$O/O$_2$ oxidation level indicating their ability to release O$_2$ in water splitting process. For conduction band edge,
(Mo, Cd) codoped SrTiO$_{3}$ lies below the H$^+$/H$_2$ reduction level but, in the case of (Mo, Zn) dopant, it is very close to the reduction level suggesting that the former may not be capable of releasing hydrogen 
in the water splitting. Hence, the codoping of (Mo, Zn) can be considered a good photocatalytic material.

\section{Conclusion} \label{conc}
 
Density functional theory is used to examine the structure, stability, electronic structure and optical properties of the co-catoinic doping of SrTiO$_{3}$ as well as those of monodopants. 
Based on the HSE hybrid exchange correlation functional, the calculated band gap of pristine SrTiO$_{3}$ is in good agreement with the experimental finding. Although the Mo and Ta monodopants decrease the band gap of the
pristine SrTiO$_{3}$, the semiconductor properties are changed which change the characteristics of the structure. The (Ta, In/Ga) codoping is energetically more 
favorable than (Mo, Zn/Cd) system. However, codoping with (Ta, In/Ga) does not improve the photocatalyitic
properties due to the larger band gap as compared to the pristine system. (Mo, Zn/Cd) codoping reduces the band gap of SrTiO$_{3}$. Since Mo and Zn comprise comparable
inoic sizes with Ti, the (Mo, Zn/Cd) co-cationic doping diminishes the band gap of pristine SrTiO$_{3}$ without emerging any undesirable mid-gap states.
(Mo, Cd) co-cationic doping has less band gap than (Mo, Zn) one but the CB edge position prevents (Mo, Cd) SrTiO$_{3}$ of being a photocatalyst for water splitting.
In this case, the bottom of the conduction band is shifted downwards while the valance band remains unaffected. Eventually, the calculations of the absorption coefficient verified the improvement of the 
(Mo, Zn) codoped SrTiO$_{3}$ water splitting photocatalytic properties.


\begin{thebibliography}{10} 

\bibitem{A1} B.\ Modak, K.\ Srinivasu, and S.\ K.\ Ghosh, RSC Adv.\ {\bf 4}, 45703 (2014).

\bibitem{A2} H.-C.\ Chen, C.-W.\ Huang, J.\ C.\ S.\ Wu, and S.-T.\ Lin, J.\ Phys.\ Chem.\ C {\bf 116}, 7897 (2012).

\bibitem{A3} B.\ Modak, K.\ Srinivasu, and S.\ K.\ Ghosh, Phys.\ Chem.\ Chem.\ Phys {\bf 16}, 24527 (2014).

\bibitem{A4} W.\ Wei, Y.\ Dai, M.\ Guo, L.\ Yu, H.\ Jin, S.\ Han, and B.\ Huang, Phys.\ Chem.\ Chem.\ Phys {\bf 12}, 7612 (2010).

\bibitem{A5} Y.\ Xu and M.\ A.\ A.\ Schoonen, Am.\ Mineral.\ {\bf 85}, 543 (2000).

\bibitem{A6} M.\ Cardona, Phys.\ Rev.\ {\bf 140}, A651 (1965).

\bibitem{A7} H.\ Yu, S.\ Ouyang, S.\ Yan, Z.\ Li, T.\ Yu and Z.\ Zou, J.\ Mater.\ Chem.\ {\bf 21}, 11347 (2011).

\bibitem{A8} D.\ Wang, J.\ Ye, T.\ Kako, and T.\ Kimura, J.\ Phys.\ Chem.\ B {\bf 110}, 15824 (2006).

\bibitem{A9} J.\ Wang, S.\ Yin, M.\ Komatsu, Q.\ Zhang, F.\ Saito and T.\ Sato, J.\ Mater.\ Chem.\, {\bf 13}, 2348 (2003).

\bibitem{A10} P.\ Liu, J.\ Nisar, B.\ Pathak and R.\ Ahuja, Int.\ J.\ Hydrogen Energy {\bf 37}, 11611 (2012).

\bibitem{A11} J.\ Wang, S.\ Yin, M.\ Komatsu, Q.\ Zhang, F.\ Saito and T.\ Sato, J.\ Photochem.\ Photobiol.\, A {\bf 165}, 149 (2004).

\bibitem{A12} C.\ Zhang, Y.\ Jia, Y.\ Jing, Y.\ Yao, J.\ Ma and J.\ Sun, Comput.\ Mater.\ Sci.\ {\bf 79}, 69 (2013).

\bibitem{A13} H.\ W.\ Kang and S.\ B.\ Park, Chem.\ Eng.\ Sci.\ {\bf 100}, 384 (2013).

\bibitem{A14} N.\ Li and K.\ L.\ Yao, AIP Adv.\ {\bf 2}, 032135 (2012).
 
\bibitem{A15} W.\ A4, Y.\ Dai, H.\ Jin, and B.\ Huang, J.\ Phys.\ D {\bf 42}, 055401 (2009).

\bibitem{A16} J.\ W.\ Liu, G.\ Chen, Z.\ H.\ Li, and Z.\ G.\ Zhang, J.\ Solid State Chem.\ {\bf 179}, 3704 (2006).

\bibitem{A17} H.\ Irie, Y.\ Maruyama and K.\ Hashimoto, J.\ Phys.\ Chem.\ C {\bf 111}, 1847 (2007).

\bibitem{A18} R.\ Konta, T.\ Ishii, H.\ Kato and A.\ Kudo, J.\ Phys.\ Chem.\ B {\bf 108}, 8992 (2004).

\bibitem{A19} S.\ W.\ Bae, P.\ H.\ Borse and J.\ S.\ Lee, Appl.\ Phys.\ Lett.\ {\bf 92}, 104107 (2008).

\bibitem{A20} Y.\ Y.\ Mi, S.\ J.\ Wang, J.\ W.\ Chai, J.\ S.\ Pan, C.\ H.\ A.\ Huan, Y.\ P.\ Feng and C.\ K.\ Ong, Appl.\ Phys.\ Lett.\
 {\bf 89}, 231922 (2006).

\bibitem{B1} G.\ Kresse and J.\ Hafner, Phys.\ Rev.\ B {\bf 47}, 558 (1993).

\bibitem{B2} G.\ Kresse and J.\ Furthmuller, Phys.\ Rev.\ B {\bf 54}, 11169 (1996).

\bibitem{B3} A.\ D.\ Becke, Phys.\ Rev.\ A {\bf 38}, 3098 (1988).

\bibitem{B4} J.\ P.\ Perdew, Phys.\ Rev.\ B {\bf 33}, 8822 (1986).


\bibitem{B5} J.\ P.\ Perdew, K.\ Burke, and M.\ Ernzerhof, Phys.\ Rev.\ Lett.\ {\bf 77}, 3865 (1996).

\bibitem{B6} H.\ J.\ Monkhorst and J.\ D.\ Pack, Phys.\ Rev.\ B: Solid State {\bf 13}, 5188 (1976).

\bibitem{B7} J.\ Heyd, G.\ E.\ Scuseria, and M.\ Ernzerhof, J.\ Chem.\ Phys.\ {\bf 118}, 8207 (2003).

\bibitem{B8} J.\ Paier, M.\ Marsman, K.\ Hummer, G.\ Kress, I.\ C.\ Gerber, and J.\ G.\ Angyan, J.\ Chem.\ Phys.\ {\bf 125}, 249901 (2006).

\bibitem{B9} J.\ Heyd, G.\ E.\ Scuseria, and M.\ Ernzerhof, J.\ Chem.\ Phys.\ {\bf 124}, 219906 (2006).

\bibitem{B10} P.\ Reunchan, N.\ Umezawa, S.\ Ouyang, and J.\ Ye, Phys.\ Chem.\ Chem.\ Phys.\ {\bf 14}, 1876 (2012).

\bibitem{B11} P.\ Reunchan, S.\ Ouyang, N.\ Umezawa, H.\ Xu, Y.\ Zhang, and J.\ Ye, J.\ Mater.\ Chem.\ A {\bf 1}, 4221 (2013).

\bibitem{C0} {\it CRC Handbook of Chemistry and Physics}, 87 th ed., edited by D.\ R.\ Lide (Taylor \& Francis, London, 2006).

\bibitem{C01} R.\ D.\ Shannon, Acta Crystallogr., Sect.\ A: Cryst.\ Phys., Diffr., Theor.\ Gen.\ Cryst.\ {\bf 32}, 751 (1976).

\bibitem{C011} Q.\ Meng, T.\ Wang, E.\ Liu, X.\ Ma, Q.\ Ge, and J.\ Gong, Phys.\ Chem.\ Chem.\ Phys.\ {\bf 15}, 9549 (2013).

\bibitem{C1} G.-Y.\ Wang, Y.\ Qin, J.\ Cheng, and Y.-j.\ Wang, Journal of Fuel Chemistry and Technology {\bf 38}, 502 (2010).

\bibitem{C2} J.-P.\ Zou, L.-Z.\ Zhang, S.\-L.\ Luo, L.-H.\ Leng, X.-B.\ Luo, M.-J.\ Zhang, Y.\ Luo, G.-C.\ Guo, International Journal of Hydrogen Energy {\bf 37}, 17068 (2012).
 
\bibitem{C03} R.\ Long and N.\ J.\ English, Chem.\ Mater.\ {\bf 22}, 1616 (2010).

\bibitem{C031} W.-J.\ Shi and S.-J.\ Xiong, Phys.\ Rev.\ B {\bf 84}, 205210 (2011).

\bibitem{C3} L.\ M.\ Sun, Y.\ Qi, C.\ J.\ Jia, and Z.\ Jin, W.\ L.\ Fan, Nanoscale {\bf 6}, 2649 (2014).

\bibitem{C4} S.\ Saha, T.\ P.\ Sinha, and A.\ Mookerjee, Phys.\ Rev.\ B {\bf 62}, 8828 (2000).



\end{thebibliography}
\end{document}